\documentclass[preprint]{elsarticle}

\usepackage{latexsym}
\usepackage{amsmath}
\usepackage{amssymb}
\usepackage{graphicx}
\usepackage{float}
\usepackage[normalem]{ulem}
\usepackage{slashed}
\usepackage{hyperref}
\usepackage{color}

\newcommand{\be}{\begin{equation}}
\newcommand{\ee}{\end{equation}}
\newcommand{\bea}{\begin{eqnarray}}
\newcommand{\eea}{\end{eqnarray}}
\newcommand{\nn}{\nonumber}

\def\Dr{\Delta_{r}}
\def\Da{\Delta_{a}}
\def\Df{\Delta_{s}}
\def\Dfb{\tilde\Delta_{s}}
\def\Dd{\Delta_{d}}
\def\Dp{\Delta_{p}}

\def\tD{\tilde\Delta}

\def\LQ{(L\cdot Q)}
\def\PQ{(P\cdot Q)}
\def\PL{(P\cdot L)}

\title{\boldmath The HTL Lagrangian at NLO: the photon case}

\author{Stefano Carignano,}
\address{Departament de F\'\i sica Qu\'antica i Astrof\'isica and Institut de Ci\`encies del Cosmos, Universitat de Barcelona, Mart\'\i$\,$ i Franqu\`es 1, 08028 Barcelona, Catalonia, Spain}
\corref{stefano.carignano@fqa.ub.edu}

\author{Margaret E. Carrington}
\address{Department of Physics, Brandon University,
Brandon, Manitoba, R7A 6A9, Canada\\
and Winnipeg Institute for Theoretical Physics, Winnipeg, Manitoba, Canada}
\corref{carrington@brandonu.ca}

\author{Joan Soto}
\address{Departament de F\'\i sica Qu\'antica i Astrof\'isica and Institut de Ci\`encies del Cosmos, Universitat de Barcelona, Mart\'\i$\,$ i Franqu\`es 1, 08028 Barcelona, Catalonia, Spain}
\corref{joan.soto@ub.edu}

\begin{document}

\begin{abstract}
We calculate the two loop hard correction to the photon self-energy  in an electron-positron plasma (EPP) for arbitrary soft momenta.  This provides the only missing ingredient to obtain the Hard Thermal Loop (HTL) effective Lagrangian at next-to-leading order (NLO), and the full photon propagator at the same order. This result can be easily extended to obtain the soft photon propagator in a quark gluon plasma.
We use the Keldysh representation of the real time formalism in the massless fermion limit, and dimensional regularization (DR) to regulate any ultraviolet (UV), infrared (IR) or collinear divergences that appear in the intermediate steps of the calculation. 
In the limit of soft photon momenta, our result is finite. It not only provides an ${\cal O}(\alpha)$ correction to the Debye mass,  
but also a new non-local structure. A consistent regularization of radial and angular integrals is crucial to get this new structure. As an application we calculate the plasmon dispersion
relations at NLO.
\end{abstract}

\maketitle
\flushbottom

\section{Introduction}
\label{sec:intro}

Originally developed in order to understand the physics of relativistic plasmas,
Thermal Field Theory is now a mature field \cite{Bellac:2011kqa,Kapusta:2006pm,Laine:2016hma}. 
 At weak coupling, for instance in QED or in QCD at high temperature
($T\gg \Lambda_{QCD}$), a number of different scales related to different physical situations
arise \cite{Arnold:1997gh}, such as the screening mass ($\sim eT$ in QED and $\sim gT$ in QCD), also called the soft scale, and the magnetic mass ($\sim g^2T$ in QCD and missing in QED), also known as the ultrasoft scale. 
If the coupling constant is small enough these scales are well separated, and effective field theory (EFT) techniques are useful in order to disentangle the physics at each scale. 

In the case of static properties (thermodynamic quantities), for QCD, the suitable EFT at the soft scale is called Electric QCD (EQCD) \cite{Ginsparg:1980ef,Appelquist:1981vg,Kajantie:1995dw}, and the EFT at the ultrasoft scale is Magnetic QCD (MQCD) \cite{Braaten:1995jr}. The latter has the peculiarity that perturbation theory in the coupling constant breaks down. For QED, at the soft scale we have an analogous EFT called EQED, but at the ultrasoft scale MQED corresponds to a free theory. These are local effective theories and the number of terms in the Lagrangian at a given order is finite and can be constructed systematically, making it possible to carry out calculations at very high order \cite{Laine:2018lgj}.

For dynamical quantities (real time phenomena), there is also an EFT suitable at the soft scale, the celebrated HTL effective Lagrangian \cite{Braaten:1991gm}. However, this effective Lagrangian is non-local and it is not clear how to construct higher order terms in the $1/T$ and weak coupling expansions, unless they are explicitly calculated. This was one of the motivations to apply the On-Shell Effective Field Theory (OSEFT) \cite{Manuel:2014dza} to this problem in ref. \cite{Manuel:2016wqs}, where the leading power corrections to the photon self-energy were calculated. The complete leading power corrections for massless QED, namely including the fermion sector, were worked out
in ref. \cite{Carignano:2017ovz}, in which the explicit form of the power suppressed terms in the Lagrangian was also given. However, this is not the full story: if the soft external momentum is of order $eT$, there can be weak coupling corrections at two loops which give rise to contributions of the same order as the leading power corrections \cite{Braaten:1989mz, Mirza:2013ula}. In this paper we calculate these contributions for the photon sector. The complete correction to the HTL Lagrangian up to NLO for this sector in $d$ space dimensions reads

\begin{align}
& {\cal L}_{HTL}^{\rm NLO}  =\frac{e^2 \nu^{3-d}}{2} \int \frac{d^d \bf p}{(2 \pi)^d} \left\lbrace \frac{ N_f(p) }{2 p^3} 
F_{\rho \alpha}   \frac{v^\alpha v^\beta}{(v \cdot \partial)^4 } \partial^4 F _{\beta}^{\,\,\rho} \right.
 \nn\\
& \left. +
e^2(d-1) \Lambda_{(d)}^2 
\left( \frac{N_f(p)}{p^3}F_{\rho\alpha} \left[\frac{v^\alpha v^\beta}{(v \cdot \partial)^2 } \left( \frac{1}{2} + \frac{\partial_0}{v\cdot \partial}\right) 
-\frac{n^\alpha v^\beta+v^\alpha n^\beta}{2(v \cdot \partial)^2 }
\right] F_\beta^{\,\,\rho} \right.\right.\nn\\
&\left.\left. -\frac{1}{2p^2}\frac{dN_f(p)}{dp}F_{\rho\alpha}\frac{v^\alpha v^\beta}{(v \cdot \partial)^2 }F_\beta^{\,\,\rho} \right)\right\rbrace
\,,
\label{1}
\end{align}
where $n^\mu =(1,{\bf 0})$, $v^\mu =(1,{\bf v})$, ${\bf v}={\bf p}/p$, $p=\vert{\bf p}\vert$ and $F_{\mu\nu}$ is the electromagnetic stress tensor. The parameter $\nu$ is the scale that is conventionally introduced in DR so that $e=e(\nu)$ remains dimensionless.
The quantities $N_f(p)$ and $\Lambda_{(d)}$ 
depend on the distribution functions and are defined later on (see (\ref{distro}) and (\ref{Delta})). 
The first term  in Eq. (\ref{1}) is of order ${\cal O}(e^2  Q^2)$ where $Q$ is the external momentum, and was obtained in \cite{Manuel:2016wqs,Carignano:2017ovz}. The second term is of order ${\cal O}(e^4 T^2 )$ and is obtained for the first time in this work. For external momentum $Q\sim e T $ both terms have the same size. 
For 
$d\to 3$ the second term is finite and the UV divergence of the first term can be removed by the usual QED counterterm \cite{Carignano:2017ovz}.

\section{Preliminaries}

In the following we will work in the Keldysh representation of the real time formalism \cite{Carrington:1997sq,Thoma:2000dc,Carrington:2006xj}.
For massless fermions and photons in the Feynman gauge, lines are defined as
\bea
iS_j(P) &= i {\slashed P}\Delta_j(P) \,,\qquad 
 -i D_j^{\mu\nu}(L) & = -i  g^{\mu\nu} \tD_j(L)\,
 \label{defs1}
\eea
where $j=r,a,p,d,s$ labels retarded, advanced, principal value, difference and symmetric propagators, respectively.
For both bosonic and fermionic fields, the symmetric propagator is obtained from the difference of the retarded and advanced propagators by multiplying by the appropriate distribution functions. Our definitions are
\begin{align}
&\Delta_{r/a}(P) = \tD_{r/a}(P) = \frac{1}{P^2 \pm ip^0 \eta}\,, \nn\\  
& \Delta_p(P) = \tD_p(P) = \frac{1}{2}\big[\Delta_r(P) + \Delta_a(P)\big] \,, \nn\\
 \Dd(P) & = \tD_d(P) = \Delta_r(P)-\Delta_a(P) 
=-2\pi i \frac{p^0}{\vert p^0\vert}\delta (P^2) \,, \nn\\
 \Delta_s(P)  &= N_f(P) \Delta_d(P) \,, \qquad \tilde{\Delta}_s(P) = N_b(P) \Delta_d(P) \,,
 \label{defs2}
 \end{align}
with $\eta\to 0^+$, $P=(p^0, {\bf p})$ and 
\bea
  N_f(P) = 1-2n_f(P) \,,\quad  N_b (P)= 1 + 2n_b (P) \,,  \\
 n_f(P) = \frac{1}{e^{p_0/T}+1} \,,\quad  n_b(P) = \frac{1}{e^{p_0/T}-1} \,, 
 \label{distro} 
\eea
where 
$T$ is the temperature.
The distribution functions fulfill 
\be
 N_b(P)  = - N_b(-P)  \,, \qquad  N_f(P) = - N_f(-P) \,, \\
\ee
and satisfy the KMS condition
\be
N_f(P_1)N_f(P_2)+N_f(P_2)N_b(P_3)+N_b(P_3)N_f(P_1)+1 = 0 \,,
\label{KMS}
\ee
which holds for arbitrary momenta  $P_1+P_2+P_3=0$.

We will calculate the retarded self-energy, from which one can reconstruct the remaining Keldysh components of the self-energy, and all components of the dressed propagator. 
We use $\Pi^{\mu\nu}(Q)$ to denote the retarded self-energy where 
$Q =(q_0,{\bf q})$, $q=\vert {\bf q}\vert$.
We will define the self-energy as $-i$ times the appropriate Feynman diagram, so that the Dyson equation gives the inverse resummed propagator as $D_{\mu\nu}^{-1}(Q) - \Pi_{\mu\nu}(Q)$, 
and the Debye mass is obtained as $m_D^2 = \lim_{{\bf q}\to 0}\Pi^{00}(q_0=0,{\bf q})$.

Our method is general and we expect it to be applicable to the study of systems that are not necessarily in thermal equilibrium.
For simplicity 
we focus on an isotropic medium in the absence of parity-violating effects. 
In this case, we can 
 decompose the photon self-energy $\Pi^{\mu\nu}(Q)$ as
\bea
\label{pi-def}
\Pi^{\mu\nu}(Q) = 
\Pi_T(n\cdot Q, Q^2) \,\bar P^{\mu\nu}  
 -\frac{Q^2}{q^2}\Pi_L(n\cdot Q, Q^2)\,P^{\mu\nu} \,,
\eea
where we have introduced 
\be
 P^{\mu\nu} =  \frac{\bar n^\mu \bar n^\nu}{\bar n^2}  \,, \qquad \bar P^{\mu\nu} = g^{\mu\nu}-\frac{Q^\mu Q^\nu}{Q^2} - P_{\mu\nu} \,,
 \ee
 with
\be
 \bar n^\mu = n^\mu - Q^\mu \,\frac{ (n \cdot Q ) }{Q^2} \,.
\ee 
 Note that $Q_\mu\bar P^{\mu\nu}=Q_\mu P^{\mu\nu}=0$, guaranteeing that the photon self-energy is transverse. The functions $\Pi_T$ and $\Pi_L$ are defined as 
 \footnote{Note that our conventions differ from those used in  \cite{Manuel:2016wqs} and \cite{Carignano:2017ovz}: more specifically, our definitions for $\Pi^{\mu\nu}$ and for $\Pi_T$ both differ by a minus sign with the ones used there. We also take this opportunity to point out some misprints in these works:
 the definition of the Debye mass after formula (45) in \cite{Manuel:2016wqs} should read $m_D^2=e^2 T^2/3$ and Eq. (45) should have the opposite sign. There are also  related misprints in  \cite{Carignano:2017ovz}: in Eq.(10) $m_D^2/2$ should read $m_D^2$, while Eq. (11) should have an overall plus sign.}
\be
   \Pi_L = \Pi^{00} 
\,, \qquad 
\Pi_T = \frac{1}{d-1}\left(\Pi^\mu_{~\mu} + \frac{Q^2}{q^2}\Pi^{00}  \right) \,,
\label{00mumu}
\ee
when working in $D=1+d$ dimensions. These expressions allow us to determine the full photon self-energy by just calculating $\Pi^{00}$ and $\Pi^\mu_{~\mu}$. 
Using the Dyson equation (defined below Eq.(\ref{KMS})) and equations
(\ref{defs1}, \ref{defs2}) the dressed propagator has the form 
\bea
D_{\mu\nu}^{\textcolor{black}{\ast}}(Q) = \frac{\bar P_{\mu\nu}}{Q^2-\Pi_T}+\frac{q^2}{Q^2}\frac{P_{\mu\nu}}{q^2+\Pi_{L}} + \frac{Q^\mu Q^\nu}{Q^4}\,.
\label{dressed}
\eea

\section{Calculation}

\begin{figure}[tbp]
\centering 
\includegraphics[scale=0.45]{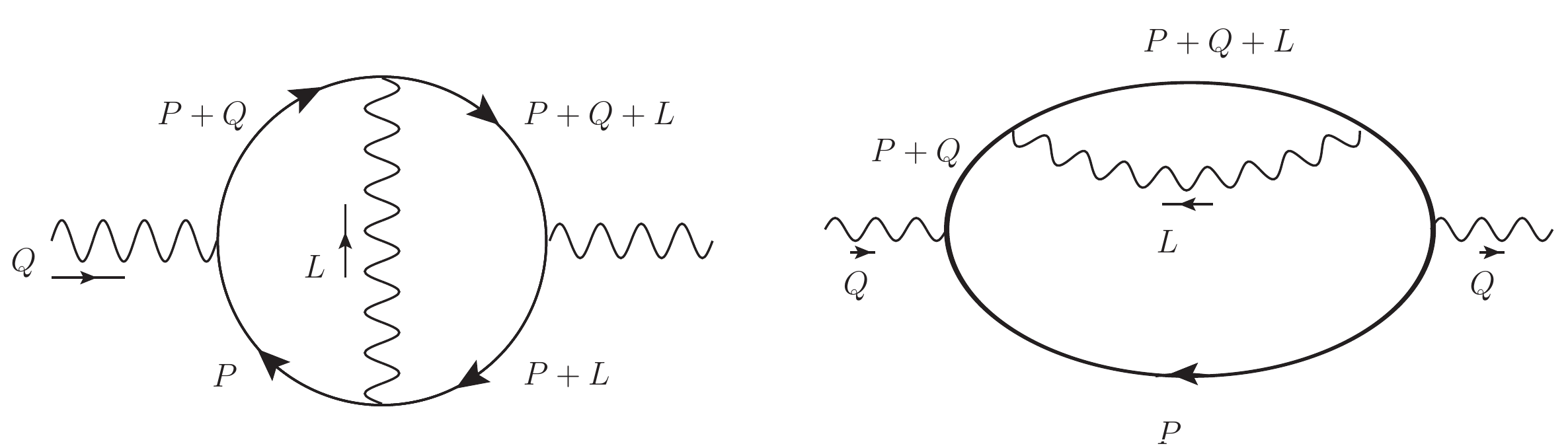}
\caption{\label{fig:1} Diagrams contributing to the photon self-energy at two loops. Left: ``vertex" diagram; right: ``self-energy" (or simply ``self") diagram.}
\end{figure}

At two loops, only two types of diagrams contribute to the photon self-energy, as shown in Fig. \ref{fig:1}. 
We call the diagrams on the left and right sides of the figure, respectively, the ``vertex'' and ``self'' diagrams.  
There is an additional diagram of the self type with the photon line starting and ending on the lower line of the loop. The contribution from this diagram is identical to the one from the self graph that is shown, and we will take it into account by including a factor of two. 

\subsection{Vertex diagram}

The vertex diagram in terms of retarded, advanced and symmetric propagators reads
(using a compact notation omitting plus signs: $LP= L+P$, $PQ = P+Q$, $LPQ = L+P+Q$)
\begin{align}
& \Pi^{\mu\nu}_{\rm (vertex)}(Q) = \frac{e^4}{4}\int \frac{d^4P}{(2\pi)^4} \int  \frac{d^4L}{(2\pi)^4} {F^{\mu\nu}_{\rm V}}(L,P,Q) \, \big\{ \label{v1} \\
& \Da(P) \Dr(L) \Da(LP) \Df(PQ) \Df(LPQ)
+\Da(P) \Dr(L) \Df(LP) \Df(PQ) \Dr(LPQ) \nonumber \\
 +&\Da(P) \Dfb(L) \Da(LP) \Dr(PQ) \Df(LPQ)
+\Df(P) \Da(L)  \Da(LP) \Dr(PQ) \Df(LPQ) \nonumber \\
+&\Da(P) \Dfb(L) \Df(LP) \Dr(PQ) \Dr(LPQ)
+\Df(P)  \Da(L) \Df(LP) \Dr(PQ) \Dr(LPQ) \nonumber \\
 +&\Da(P) \Dfb(L) \Da(LP) \Df(PQ)\Da(LPQ)
+\Da(P) \Dr(L) \Da(LP) \Dr(PQ) \Da(LPQ)\nonumber \\
+& \Df(P) \Da(L)  \Dr(LP)\Dr(PQ) \Dr(LPQ)
+ \Df(P) \Dfb(L)  \Dr(LP) \Dr(PQ) \Dr(LPQ)\nonumber
   \big\}\,,
\end{align}
where we have dropped terms that depend on 
propagators that are all advanced, or all retarded, with respect to the variable $L$,
as they vanish after contour integration.
The Dirac structure of the fermion propagators and vertices can be factored out in the  trace
\be
{F^{\mu\nu}_{\rm V}} = \text{Tr}\big[\gamma^\mu \; \slashed{P}\;\gamma_\lambda \; (\slashed{P}+\slashed{L})\;\gamma^\nu \;(\slashed{P
}+\slashed{L}+\slashed{Q})\; \gamma^\lambda \;(\slashed{P}+ \slashed{Q}) \big]\,.\nonumber
\label{trvertex}
\ee
The diagonal elements of this trace ($\mu=\nu$), which are the only ones needed for our calculation, are invariant under the following changes of variables
\bea
\label{ch1} & (a) &  P\to -P-L-Q \,, \\ 
\label{ch2} & (b) & P\to -P-Q\,,\quad L\to -L \,.
\eea
These changes of variables, together with the KMS relation,
allow to reduce (\ref{v1}) to the following three terms
(here and in the following the notation $ \Pi^{\mu\mu}$ denotes equal indices and summation is implied only if the two indices appear one up and one down)
\begin{align}
 \Pi_{\rm (vertex)}^{\mu\mu}(Q) & = \frac{e^4}{2}\int \frac{d^4P}{(2\pi)^4} \int  \frac{d^4L}{(2\pi)^4} {F^{\mu\mu}_{\rm V}}(L,P,Q) N_f(P) \nonumber \\
& \Big\lbrace  N_f(Q+P+L) \Da(L) \Dd(P) \Dr(P+Q) \Da(L+P) \Dd(Q+P+L) \nn\\
& +2 N_b(L) \big[ \Dd(P) \Dd(L) \Dr(P+Q) \Dp(P+L) \Dr(Q+P+L) \nn\\ &+ \Dp(L) \Dd(P)  \Dr(P+Q) \Dd(P+L)  \Dr(Q+P+L) \big] \Big\rbrace \,,
\label{v2}
\end{align}
where we have dropped terms which do not depend on the distribution functions and therefore do not contain the medium effects we are interested in.
Finally, we use KMS (\ref{KMS}) together with some additional variable changes (which need not leave the trace invariant) to bring our expression into a form 
where all terms contain the factor $\Delta_d(P)\Delta_d(L)$, which will be most convenient when performing the integrations. 
In particular, we will employ 
\bea
\label{ch3} & (c) &  L\to -P-L-Q \,, \\ 
\label{ch4} & (d) & L \to  L-P \,.
\eea
 We arrive at
\begin{align}
& \Pi^{\mu\mu}_{\rm (vertex)}(Q) = \frac{e^4}{2}\int \frac{d^4P}{(2\pi)^4} \int  \frac{d^4L}{(2\pi)^4} N_f(P) \Dd(P) \Dd(L) \nn\\
\Big[& N_f(L) \Dr(P+Q) \Dr(L+Q) \Dr(P+Q+L) F^{\mu\mu}_{\rm V (c)} \nn\\
+& 2 N_b(L) \Dr(P+Q) \Dp(L+P) \Dr(P+Q+L) F^{\mu\mu}_{\rm V} \nn\\
+& N_f(L) \Dp(L-P)\Dr(P+Q)\Dr(L+Q) F^{\mu\mu}_{\rm V (d)} \Big] \,,
\label{eq:finVertex}
\end{align}
where we labeled by $F_{(i)}$ the result of the trace after a given variable change, so in our notation $F^{\mu\mu}_{V(a)} = F^{\mu\mu}_{V(b)} = F^{\mu\mu}_V$.

\subsection{Self-energy diagram}

The expression for the second diagram of Fig. \ref{fig:1} in terms of retarded, advanced and symmetric propagators reads, after dropping again medium-independent terms,
(again in compact notation)
\begin{align}
& \Pi^{\mu\nu}_{\rm (self)}(Q) =  \frac{e^4}{4}\int \frac{d^4P}{(2\pi)^4} \int  \frac{d^4L}{(2\pi)^4}  F^{\mu\nu}_{\rm S}(L,P,Q) \Big\lbrace \nn \\ 
& \Dfb(L)\Df(P)\Dr(PQ)\Dr(PQ)\Dr(LPQ) 
+ \Dfb(L)\Da(P)\Df(PQ)\Dr(PQ)\Dr(LPQ) \nn\\
 + & \Dfb(L)\Da(P)\Df(PQ)\Da(PQ)\Da(LPQ)
+ \Dfb(L)\Da(P)\Dr(PQ)\Da(PQ)\Df(LPQ) \nn\\
+ & \Da(L)\Df(P)\Dr(PQ)\Dr(PQ)\Df(LPQ) 
 + \Da(L)\Da(P)\Df(PQ)\Dr(PQ)\Df(LPQ) \nn \\
+ & \Dr(L)\Da(P)\Df(PQ)\Da(PQ)\Df(LPQ) \Big\rbrace \,.
\label{eq:piself1}
\end{align}
The Dirac trace in this case is given by
\be
F^{\mu\nu}_{\rm S} = \text{Tr}\big[\gamma^\mu \; \slashed{P}\;\gamma^\nu (\slashed{P}+\slashed{Q}) \gamma_\lambda \;(\slashed{Q}+\slashed{P}+\slashed{L})\;\gamma^\lambda \;(\slashed{P} +\slashed{Q})\;  \big] \,.\label{self-fac3} \nonumber
\ee
The expression in Eq. (\ref{eq:piself1}) appears to contain pinch singularities, but they can be shown to vanish using the KMS
 relation (\ref{KMS}).
 We are then left with the following terms,
\begin{align}
 \Pi^{\mu\nu}_{\rm (self)}(Q) = & \frac{e^4}{4}\int \frac{d^4P}{(2\pi)^4} \int  \frac{d^4L}{(2\pi)^4}  F^{\mu\nu}_{\rm S}(L,P,Q) \Big\lbrace \nn \\ 
 & \Df(P) \Dr(PQ)^2 \left(\Da(L) \Df(LPQ)+\Dfb(L) \Dr(LPQ)\right) \nonumber\\
 + &\Da(P) \Dfb(L) N_f(PQ) \left[\Dr(PQ){}^2 \Dr(LPQ)-\Da(PQ){}^2 \Da(LPQ)\right] \nonumber\\
+ & \Da(P) N_f(PQ) \Df(LPQ) \left[\Da(L) \Dr(PQ){}^2-\Dr(L) \Da(PQ){}^2\right] \big.    \Big\rbrace\,.
\end{align}
As for the vertex diagram,
we now make changes of variables to bring our expression to a simpler form in which the occupation numbers depend on individual internal loop momenta.
In this case however none of the variable changes will leave the trace invariant so that all combinations need to be computed separately.
A recurring combination given by the difference of the squares of two propagators can be conveniently rewritten as follows:
\be
\Delta_r^2(P)-\Delta_a^2(P)=\left.\frac{d}{dM^2}\Big[\Delta_r(P,M)-\Delta_a(P,M)\Big]\right\vert_{M^2=0}=\left.
\frac{d}{dM^2}\Delta_d(P,M)\right\vert_{M^2=0}\,,
\ee
where we have introduced $\Delta_j(P,M) \equiv \Delta_j(P)\vert_{{\bf p}^2  \rightarrow {\bf p}^2 + M^2} $.

After these simplifications only five terms survive and we arrive at 
\begin{align}
 \Pi^{\mu\nu}_{\rm (self)}(Q) &= \frac{e^4}{4}\int \frac{d^4P}{(2\pi)^4} \int  \frac{d^4L}{(2\pi)^4}  N_f(P) \Dd(L) \nn\\
\Big\lbrace
& \Da(P+Q)^2 \Dd(P)  \Da(Q+P+L) \left[ N_b(L) F^{\mu\nu}_{\rm S}+ N_f(L) F^{\mu\nu}_{\rm S(c)} \right] \nn\\
+& 
\Dr(P+L) \frac{d}{dM^2} \Dd(P,M) \Dr(P+Q) \left[ N_b(L)F^{\mu\nu}_{\rm S(b)}  + N_f(L) F^{\mu\nu}_{\rm S(e)} \right] \nn\\
+&
 \Da(P+L)^2 \Dd(P)  \Da(Q+P+L)  N_b(L) F^{\mu\nu}_{\rm S(a)} 
 \Big\rbrace \,,
\label{eq:finSelf}
\end{align}
where we introduced a fifth variable change $(e): L \rightarrow L+P \,, P \rightarrow -P-Q$,
and the $M^2$ derivative will be taken after we have performed the $p^0$ integral using the delta function in the factor $\Delta_d(P,M)$. 

\subsection{$\Pi^\mu_{~\mu} (Q)$}

\label{trace-component}

The full result for $\Pi^\mu_{~\mu} (Q)$ is obtained by combining $\Pi^\mu_{~\mu \rm (vertex)} + 2 \Pi^{\mu}_{~\mu\rm (self)}$. By inspecting the momentum structure of our expressions, we observe that for arbitrary $L$ there are collinear divergences when $L$ is parallel to $P$, hence the expressions obtained so far are formal and must be regulated. A convenient way to do this 
is to use dimensional regularization (DR). 
In certain kinematical limits the collinear divergences cancel exactly. This is the case in the soft $L$ limit, as considered in ref. \cite{Aurenche:2003ac}, and also in the case we are interested in, namely the soft $Q$ limit.
In order to be consistent within our DR prescription we also check for possible additional contributions from performing the Dirac traces in $D=d+1=4+2\epsilon$ dimensions. The evaluation of $F_V$ and $F_S$ thus leads to additional pieces coming from the relation $\gamma^\mu\gamma_\mu = g^\mu_{~\mu} = D$. In the following we will focus on the small $\epsilon$ limit, even though we will later provide our final result for arbitrary $D$. \\

Expanding for small $Q$, regularizing with DR and exploiting the symmetry of the integrand when possible to interchange $L$ with $P$, we obtain \footnote{In the denominators below, we have dropped the $\pm i\eta$ prescriptions. This can safely be done if we assume $q^0> q$, or equivalently $Q^2>0$. The arbitrary case can be recovered by replacing $q^0\to q^0+i\eta$.}
\begin{align}
\Pi^{\mu}_{\mu}(Q)  & = 
- 2 e^4 \nu^{6-2d}\int d l^0 \int d p^0 \int\frac{d^d \bf l}{(2\pi)^d}\int\frac{d^d\bf p}{(2\pi)^d}  
\,\delta(L^2) \delta(P^2)  \left[ N_b(L) - N_f(L) \right]   \nn\\ 
 \Bigg\lbrace &  N_f(P) (1+\epsilon) \left[ \frac{3}{2} \frac{Q^2}{\LQ \PQ} - \frac{2}{\PQ} + \frac{Q^2}{\PQ^2}  - \frac{2}{\PL} \right] \nn\\
&- \frac{1}{p} (1+2\epsilon) \left[ \frac{dN_f(P)}{d{p^0}} - \frac{N_f(P)}{{ p^0}} \right]
+ {\cal O}(\epsilon^2)
 \Bigg\rbrace \,.
 \label{eq:pimumusmallQ1}
\end{align}
After doing the $p^0$ and $l^0$ integrals by employing the delta functions, several of these pieces (including those containing a collinear divergency $\sim 1/\PL$ for any $\pm i\eta$ prescription) vanish due to the symmetry of the integrand.  In principle ${\cal O}(\epsilon^2)$ pieces from the trace might also contribute if both a radial and a collinear divergence are present. We however checked separately that these pieces are also collinear-safe, so that they do not contribute, and they have thus been omitted in our expression.  
Note that (\ref{eq:pimumusmallQ1}) holds for any distribution functions that fulfill the conditions (\ref{KMS}).

Exploiting the parity invariance of the distribution functions, we can now introduce the four-velocities associated with the on-shell positive and negative-energy massless degrees of freedom involved in our calculation and write $L^\mu = \pm l v_l^\mu$, $P^\mu = \pm p v^\mu$, with $v_l^\mu = (1, {\bf v}_l)$ and $v^\mu = (1, {\bf v})$ (see e.g. \cite{Carignano:2017ovz} for details).
 The angular part of the integral  over ${\bf l}$ is trivial and the radial integral over $|{\bf l}|=l$  depends exclusively on the combination $N_f(l) - N_b(l)$, and gives a finite result. The remaining integral over ${\bf p}$ requires however a bit more attention.
Focusing again on the $\epsilon\to 0$ limit, the relevant contributions are 

\begin{align}
\Pi^{\mu}_{~\mu}(Q) & =  - 2 e^4 \nu^{3-d} \Lambda_{(3)}^2 \int \frac{d^d\bf p}{(2\pi)^d} \Bigg\lbrace  \frac{1}{p^2}\frac{dN_f(p)}{dp} \nn\\
& +   \frac{N_f(p)}{p^3} \Big[  \left(\frac{Q^2}{(v\cdot Q)^2} - 1 \right) 
  + \epsilon  \left( \frac{Q^2}{(v\cdot Q)^2} - 2 \right)   \Big] \Bigg\rbrace \,,  
  \label{eq:pimumusmallQ2}
\end{align}
where we have introduced the quantity
\be 
\Lambda_{(d)}^2=\nu^{3-d}\int \frac{d^d \bf l}{(2\pi)^d} \left( \frac{ N_b(l) - N_f(l) }{l} \right) \,.
\label{Delta}
\ee
For thermal distributions, 
$\Lambda^2_{(3)}=T^2/4$.
The radial $p$ integral in the second line of (\ref{eq:pimumusmallQ2}) is formally divergent when $d\to3$, whereas in the same limit the angular integral of the first term in the square bracket vanishes. Thus we see that some additional care is required when considering this contribution.
After a proper treatment of these two integrals in DR (see e.g. \cite{Carignano:2017ovz} for a collection of the relevant formulas), we find that the radial integral is ${\cal O} (1/\epsilon)$, whereas the angular integral over the first term in the square bracket is ${\cal O} (\epsilon)$, so that the final result for their product is finite. The angular integral over the second term in the square brackets is finite and hence, due to the explicit 
factor of $\epsilon$ in front of the round bracket (which was produced by the $D=4+2\epsilon$ dimensional trace), this second term also
gives a finite result.
Including the finite contribution from the first line of (\ref{eq:pimumusmallQ2}), we arrive at 
\be
\Pi^{\mu}_{~\mu}(Q) = -\frac{e^4T^2}{8\pi^2} \left[ 1 + \frac{q^0}{q} \log\left(\frac{q^0+q + i\eta}{q^0-q+i \eta} \right) \right]\,,
\label{mumu}
\ee
where we have restored the $+i\eta$ prescription for retarded quantities.
We would like to stress here that the use of a consistent regularization scheme is crucial in order to obtain this result: in particular, 
if one works in $D=4$ and regulates the radial integral with a naive cut-off, 
one misses the contribution of the second line of (\ref{eq:pimumusmallQ2}), and hence the second term in (\ref{mumu}).

\subsection{$\Pi^{00} (Q)$}

The full result for $\Pi^{00} (Q)$ is obtained by adding $\Pi^{00}_{\rm (vertex)} + 2 \Pi^{00}_{\rm (self)}$.
In this case expressions are lengthier, but upon carrying out the small $Q$ expansion at leading order and performing the $l^0$ and $p^0$ integrations, 
we again find that all collinear divergences cancel and arrive at {\footnotemark[2]}
\begin{align}
\Pi^{00}(Q) & =   e^4 \nu^{3-d}\Lambda_{(d)}^2 
\int \frac{d^d{\bf p}}{(2\pi)^d}  \Bigg\lbrace \frac{N_f(p)}{p^3} \left[ 1 + \frac{2 q^0 Q^2}{(v \cdot Q)^3} - \frac{Q^2}{(v \cdot Q)^2} - 2 \frac{q_0^2}{(v\cdot Q)^2}  \right] \nn\\
&- \frac{1}{p^2} \frac{dN_f(p)}{dp} \left[ 1 + \frac{Q^2}{(v\cdot Q)^2} - 2 \frac{q^0}{(v\cdot Q)} \right] \Bigg\rbrace + {\cal O}(\epsilon) \,.
\label{eq:pi00smallQ2}
\end{align}
As in section \ref{trace-component}, for $d=3$ we find a diverging radial integral multiplied by a vanishing angular one.
 Using DR we find again that the divergent part of the radial integral $(\sim 1/\epsilon)$ and the vanishing angular one $(\sim \epsilon)$ combine into a finite expression,
and we arrive at
 \be
\Pi^{00}(Q) =  - \frac{e^4T^2}{8\pi^2} \left[ 1 - \frac{q_0^2}{Q^2} \right]\,.
\label{pi00}
\ee

As {in the case of} the $\Pi^{\mu}_{~\mu}$ component, we have checked for additional possible contributions stemming from performing the trace in $D$ dimensions. In this case only ${\cal O}(\epsilon)$ pieces are present, and - unlike the previous case - we find them to be exactly proportional to the result for $D=4$, which we showed above to be finite. We can thus safely take the $\epsilon\to 0$ limit and see that they do not contribute.

As a cross-check of our result, we can compute the limit
\be
 \lim_{{\bf q}\to 0} \Pi^{00}(q^0=0,{\bf q}) = -\frac{e^4T^2}{8\pi^2} \,,
\ee
which is consistent with results in the literature and can be associated with a NLO correction to the Debye mass \cite{Kapusta:2006pm}.
Once again, if one works in $D=4$ and naively regulates the radial integrals with a hard cut-off, one misses the contribution from the first line of (\ref{eq:pi00smallQ2}),
leading to an incorrect result in (\ref{pi00}) (specifically, $\Pi^{00}(Q)$ would be proportional to the HTL result).

\section{Results and outlook}

In this work, we computed the two-loop hard contribution to the photon self-energy in a medium at leading order for small momentum using the real-time formalism. This provides the missing ingredient 
to build the full NLO contribution to the HTL effective action. 
The contribution is found to be finite for $d=3$ as a result of a non-trivial series of cancellations of divergences which we treated consistently using dimensional regularization. 

From the computed results for $\Pi^{00}$ and $\Pi^{\mu}_{~\mu}$ we can reconstruct the photon self-energy associated with this contribution using Eqs.  (\ref{pi-def}) and (\ref{00mumu}). However, in order to find  the
effective Lagrangian from which these contributions are derived, it is convenient to reconstruct $\Pi^{\mu\nu}$ before the last angular integral is performed. It is straightforward to check that (\ref{mumu}) and (\ref{pi00})   can be obtained from,
\be
\Pi^{\mu\nu}(Q)= -\frac{e^4T^2}{8\pi^2} \int \frac{d\Omega_v}{4\pi} \Big( \frac{1}{2} + \frac{q^0}{v\cdot Q} \Big) A^ {\mu\nu}
\,, 
\ee
where
\be
A^{\mu\nu}  = v^\mu v^\nu \frac{Q^2}{(v\cdot Q)^2} - \frac{v^\mu Q^\nu + v^\nu Q^\mu}{v\cdot Q} + g^{\mu\nu}  \, 
\ee
is the same structure that appears in the HTL self-energy. 
Hence, the corresponding effective Lagrangian reads
\be
{\cal L} = - \frac{e^4T^2}{16\pi^2}\int \frac{d\Omega_v}{4\pi}F_{\rho\mu}\frac{v^\mu v^\nu}{(v \cdot \partial)^2 } \left( \frac{1}{2} + \frac{\partial_0}{v\cdot \partial}\right) F_\nu^{\,\,\rho}\,.
\ee

For completeness, we also give the result in arbitrary dimension $D=d+1$, not necessarily close 
to the $D=4$ case. 
For this we need an extra structure,
\begin{align}
B^{\mu\nu} & = \frac{n^\mu Q^\nu + n^\nu Q^\mu}{v\cdot Q} + \frac{v^\mu Q^\nu + v^\nu Q^\mu}{v\cdot Q} \frac{q^0}{v\cdot Q} - (n^\mu v^\nu + n^\nu v^\mu)  \frac{Q^2}{(v\cdot Q)^2} - g^{\mu\nu} \frac{2q^0}{v\cdot Q} \,. 
\end{align}
We can then write, 
\begin{align}
\Pi^{\mu\nu}(Q) &= 
e^4 \nu^{3-d} \Lambda_{(d)}^2\Big(\frac{d-1}{2}\Big)\int\frac{d^d\bf p}{(2\pi)^d}\Bigg\lbrace \frac{N_f(p)}{p^3} \Big[ \Big( 1  + \frac{2 q^0}{v\cdot Q} \Big) A^{\mu\nu} \nn\\
& + B^{\mu\nu} \Big] - \frac{1}{p^2}\frac{dN_f}{dp} A^{\mu\nu} \Bigg\rbrace \,, 
\end{align}
which trivially fulfills the Ward identity $Q_\mu \Pi^{\mu\nu}(Q)= 0$, since both $A^{\mu\nu}$ and $B^{\mu\nu}$ are transverse. 
This self-energy agrees with the result in (\ref{eq:pimumusmallQ2}) when $\epsilon\to 0$ (as explained in the paragraph under equation (\ref{Delta})) and can be obtained from the second term in (\ref{1}), which verifies the claim that (\ref{1})
is indeed the NLO Lagrangian of the HTL effective theory. 

The Lagrangian (\ref{1}) describes the dynamics of soft photons in a medium at NLO for arbitrary kinematics. 
With minor modifications,
it also describes the NLO dynamics of soft photons in a QCD medium: in the first term $e^2\to \sum_{q=u,d,s} Q_q^2 e^2$ and in the second term $e^4\to \sum_{q=u,d,s} Q_q^2 e^2 C_F N_c g^2$, where $Q_q$ is the charge of the quark $q$, $g$ is the QCD coupling constant, $N_c=3$ is the number of colors, and $C_F=(N_c^2-1)/2N_c$.

We emphasize that the two-loop self-energies calculated in this paper, together with the power corrections to the one-loop self-energies calculated in \cite{Carignano:2017ovz}, provide the full NLO soft photon propagator for both an EPP and a QGP.
This result follows from the fact that:
a) for the two-loop diagrams, a soft internal photon is suppressed by a factor of $e$ (or $g$), and a soft internal fermion is even more suppressed; and 
b) the contribution of soft fermions to the one-loop soft photon self-energy is also suppressed (${\cal O}(e^5 T^2)$ in an EPP and ${\cal O}(e^2 g^3 T^2)$ in a QGP) \cite{Mirza:2013ula}.
Our result can therefore be applied to any e.m. process involving soft photons
for which the precision needed includes ${\cal O}(e^2, Q^2/T^2)$ corrections.
As an example, we calculate the dispersion relation at NLO in \ref{app:b}.

\appendix

%
\section{Plasma oscillations}
\label{app:b}

In this Appendix we present NLO results for the plasma frequencies of photonic collective modes. 
We remind the reader these dispersion relations can be obtained from the two-loop self-energies calculated in this paper, together with the power corrections to the one-loop self-energies calculated in \cite{Carignano:2017ovz}, because two-loop diagrams with a soft line and one-loop diagrams with a soft fermion are relatively suppressed \cite{Mirza:2013ula}. 
Notice
that this is different from the non-abelian case, where extra soft
contributions would modify the result.

We solve the dispersion equations obtained from the poles of Eq. (\ref{dressed}):
\bea
Q^2-\Pi_T(q^0,q)=0\,,\quad q^2 + \Pi_L(q^0,q)=0\,.
\label{disp}
\eea
 Both $\Pi_L$ and $\Pi_T$ contain three different contributions: the LO HTL result, the power corrections to the 1-loop HTL result (calculated in ref. \cite{Manuel:2016wqs}), and the 2-loop result calculated in this paper. For convenience we gather these results in Eq. \ref{fullPi}. 
 \bea
 &&\Pi_L^{\rm htl} = \frac{e^2T^2}{3}\left(1- \frac{q_0}{2q} \log \left(\frac{q+q_0}{q_0-q}\right)\right)\nn\\
&& \Pi_L^{\rm pow\cdot corr} = -\frac{e^2}{4 \pi ^2}
 \Big(q^2  -\frac{q_0^2}{3}\Big)\Big( 1 - \frac{q_0}{2q} \log\left(\frac{q+q_0}{q_0-q}\right)\Big)\nn\\
&& \Pi_L^{\rm 2loop} = \frac{e^4 T^2 q^2}{8 \pi ^2 Q^2}\nn\\
&&  \Pi_T^{\rm htl}= \frac{e^2T^2}{3}\frac{q_0}{4 q^3} \left(2 q q_0-Q^2 \log \left(\frac{q+q_0}{q_0-q}\right)\right)\nn\\
&& \Pi_T^{\rm pow\cdot corr} = \frac{e^2}{4 \pi ^2} \left(\frac{q_0^2}{2}+\frac{q_0^4}{6 q^2}
-\frac{2 q^2}{3}
-\frac{q_0^3}{12 q^3}  \left(2 q^2+q_0^2 -\frac{3 q^4}{q_0^2}\right)\log \left(\frac{q+q_0}{q_0-q}\right)  
\right)\nn\\
&& \Pi_T^{\rm 2loop} = -\frac{e^4 q_0 T^2}{16 \pi ^2 q} \log \left(\frac{q+q_0}{q_0-q}\right) \,.
\label{fullPi}
 \eea
For simplicity, we have taken $\Pi_I^{\rm pow\cdot corr}$ , $I=L,T$, at the scale $\nu=Te^{-\gamma_E/2-1}\sqrt{\pi}/2$ in the MS scheme. This fixes the scale of $e^2=e^2(\nu)$ in $\Pi_I^{\rm htl}$. The screening mass, defined as $m_S^2=\Pi_L(0,q)\big|_{q^2=-m_S^2}$, calculated from these results reads
\be
m_S^2=T^2\left(\frac{e^2}{3}-\frac{e^4}{24\pi^2}\right)\,
\ee
and agrees with the results of Ref. \cite{Blaizot:1995kg} at NLO.\footnote{The agreement is also mantained if the scale $\nu$ is left arbitrary.}

\begin{figure}[H]
\begin{center}
\begin{minipage}{6cm}
\includegraphics[width=5.6cm]{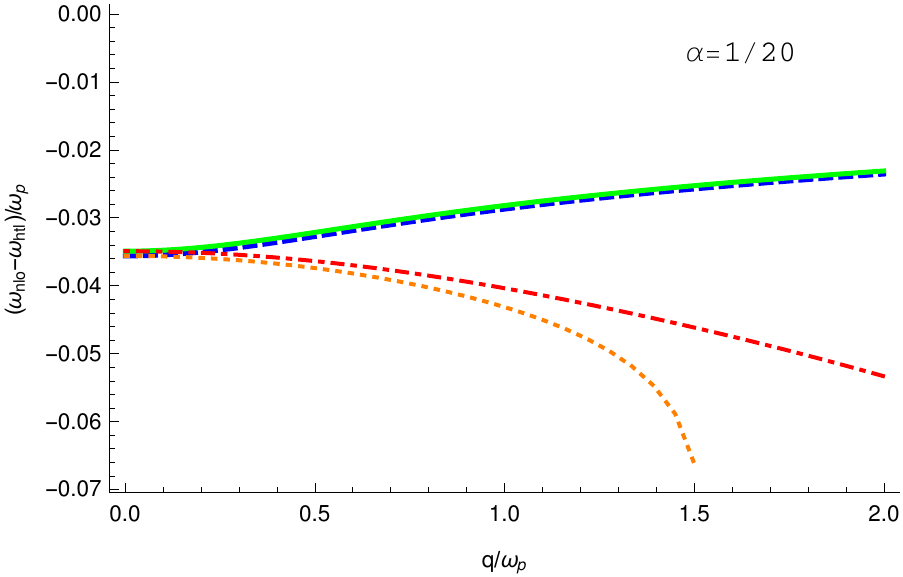}
\end{minipage}
\begin{minipage}{6cm}
\includegraphics[width=5.6cm]{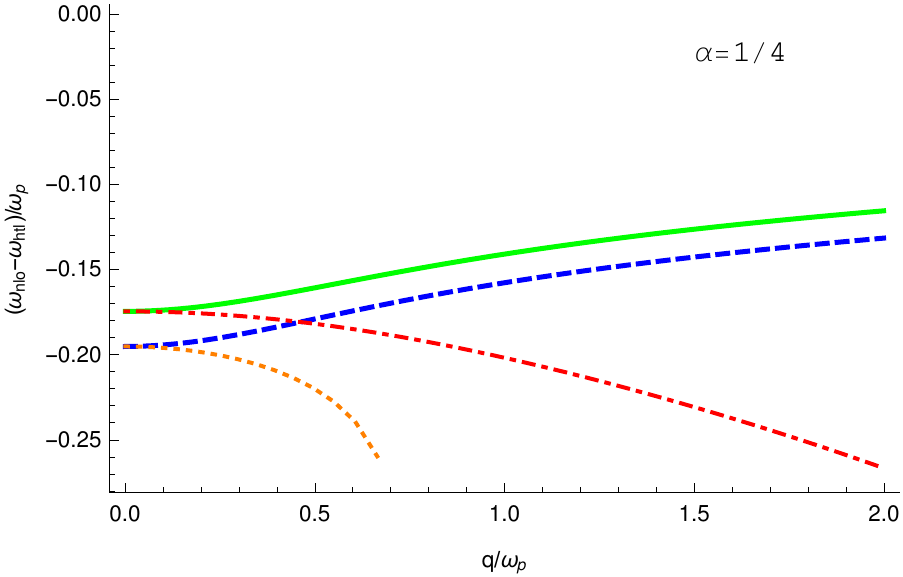}
\end{minipage}
\end{center}
\caption{Next-to-leading order corrections to the HTL dispersion relations for the full transverse (blue, dashed), expanded transverse (green, solid), full longitudinal (orange, dotted) and expanded longitudinal (red, dot-dashed) modes, for two different values of $\alpha$.
For high values of q the NLO corrections generate imaginary parts in the longitudinal dispersion
relations and the solutions become complex - we stop plotting our results when this occurs.
  }
 \label{omega-nlo}
\end{figure}

We substitute the expressions in (\ref{fullPi}) into the dispersion equations (\ref{disp}), $\Pi_I= \Pi_I^{\rm htl}+\Pi_I^{\rm pow\cdot corr}+\Pi_I^{\rm 2loop}$, $I=L,T$, scale all dimensional quantities by $\omega_p=eT/3$,
and solve the resulting equations numerically for $q^0=\omega_{\rm nlo}$. 
To obtain an estimate of the range over which the solutions are valid, we have also solved the equations within a consistent perturbative expansion. To do this we write the dispersion equations in the form
\bea
 && q^2 = -
\big[\Pi_L^{\rm htl}(\omega_{\rm{htl}} + \delta,q)  + \Pi_L^{\rm pow\cdot corr}(\omega_{\rm{htl}},q) + \Pi_L^{\rm 2loop}(\omega_{\rm{htl}},q)\big] \,, \nonumber\\
&& (\omega_{\rm{htl}} + \delta)^2-q^2 = \big[\Pi_T^{\rm htl}(\omega_{\rm{htl}} + \delta,q)  + \Pi_T^{\rm pow\cdot corr}(\omega_{\rm{htl}},q) + \Pi_T^{\rm 2loop}(\omega_{\rm{htl}},q)\big]\,. \nonumber\\
\label{D-expanded}\eea
In each dispersion equation, $\omega_{\rm htl}$ is the well known HTL solution which is obtained when the power correction and 2 loop correction terms are dropped. We use the definition $\delta = \omega_{\rm nlo}-\omega_{\rm htl}$ to represent the difference between the next-to-leading order solution and the leading order HTL solution. We obtain perturbatively consistent solutions by expanding the equations in (\ref{D-expanded}) to linear order in $\delta$.
In Fig. \ref{omega-nlo} we show the next-to-leading order plasma frequencies relative to the leading order HTL results, using both the full expressions for the dispersion equations and the consistently expanded versions, and for two different values of $\alpha=e^2/(4\pi)$. We note that the plasma frequencies of longitudinal and transverse modes at $q=0$ coincide at NLO, as they do in the HTL approximation. 
\par

\section*{Acknowledgments}

We thank Cristina Manuel for collaboration in the early stages of this project, and for  critical reading of the manuscript. SC and JS acknowledge financial support from  the FPA2016-76005-C2-1-P project (Spain), and the 2017-SGR-929 grant (Catalonia). JS has also been supported by the FPA2016-81114-P project (Spain). MEC thanks the Institut de Ci\`encies del Cosmos, Universitat de Barcelona for hospitality during the initial phase of this work, and acknowledges support from the Natural Sciences and Engineering Research Council of Canada.

\end{document}